# Mathematical Demonstration Darwinian Theory of Evolution

Vernon Williams–vernvlw81@gmail.com

**Abstract** Darwin's book, *Origin of the Species* has been a source of public controversy for more than hundred and fifty years. Court cases and mountains of words have not dispelled this controversy. In this paper, a quantitative approach using simple mathematics shows that the concept of evolution by natural selection using only random choice of variables does work. The procedure applied to the optical equations forming the phenotype of a spider eye produces an eye design modeled after the measurements by Land and Barth.

## 1. INTRODUCTION

As Feynmen has said, "People who wish to analyze nature without using mathematics must settle for reduced understanding." This paper is an attempt to aid in this regard.

Darwin pondered whether his evolution theory and its principle of random choice could explain the development of precise organs like the eye, which require accuracy sometimes to a portion of a light wavelength like 0.000005 parts of an inch. Darwin, in his *Origin of the Species* [Dar 1859], expressed his thinking as follows:

> *To suppose that the eye, with all its inimitable contrivances for adjusting the focus to different distances, for admitting different amounts of light and for the correction of spherical and chromatic aberration, could have been formed by natural selection, seems I freely confess, absurd in the highest possible degree. Yet reason tells me, that if numerous graduations from a perfect and complex eye to one very imperfect and simple, each grade being useful to possessor, can be shown to exist; if further, the eye does vary ever so slightly, and the variations be inherited, which is certainly the case; and if any variation or modification in the organ be ever useful to an animal under changing conditions of life, then the difficulty of believing that a perfect and complex eye could be* *formed by natural selection, though insuperable by our imagination, can hardly be considered real.*

This paper examines the phenotype of the PM eye of the ctenid spider, *cupiennius sale*, to evaluate mathematically whether Darwin's theory of evolution holds up. The procedure is to simulate eye evolutions using a computer program that cycles optical equations that define the eye. The variables for the equations are chosen by random means. This procedure correctly defines the structure of the eye no matter its degree of completion. The type of eye examined is a spider eye. Although this procedure examines only a mutated eye, a part of spider evolution as a whole, this method of examining a part of the whole is the essence of Western scientific thinking [Bar 1991].

## 2. OVERVIEW OF COMPUTER CALCULATIONS

According to Atmar [Atm 1997], Darwin's theory of evolution requires five components: 1) a bounded arena, 2) a replicating population that must eventually expand beyond the bounds of the arena, 3) thermodynamically inescapable replicate error, 4) competition for space in that arena among inevitable variants, and 5) consequential competitive exclusion of the lesser fit. A PC computer cycles thousands of times to simulate the many generations of spiders where evolution takes place.

The ideal way to examine the evolution of the eye would be to evaluate how the genotype defines the formation of the eye. However, the knowledge of the genotype in regards the eye is still unknown except for fragments of information [Ghe2005]. The optical equations for the PM eye contain variables that are changed randomly. Thes equations define the phenotype. However, but the genotype resides in the phenotype [Lang 1987]. Thus, if a phenotype is removed from the gene pool so is the genotype. In Darwin's time, only the phenotype was available so it is believed that manipulating the phenotype is a valid technique.

A spider with a mutant eye lens, a rare event, but it must compete for food and reproduction rights with its siblings and parents that have do not have mutated eyes. Darwin believed this competition was the source of natural selection. In this paper, the metric for natural selection for spider with mutated eyes is an f-number (f/#) in the range of 0.55 to 0.85 (extremely small numbers) and a SS, imaged spot size, less than 0.006 mm. The reasons for these choices are that the small f/# implies the spider can hunt prey in dim light and SS< 0.006 mm implies excellent acuity for the mutated eye. Spiders that do not meet this condition are removed.

According to Darwin, mating is a lesser form of natural selection and is not as significant as natural selection [Dar 1859, p 136]. Moreover, Mendel [Men 1886] also showed that offspring inherit traits from each parent by a probability function that is difficult to specify except through statistics. Because of the difficulty of including both these effects, they were not included in this attempt to demonstrate evolution by simple mathematics..

In general, animal eyes do not have exactly spherical lens surfaces because a spherical lens will contain spherical aberration that distorts the image. Conic correction for the spherical surface is an easy method to correct this aberration. Nature demonstrates this principal in the trilobite. In the last group of trilobites after evolving over a millions of years and before extinction, these animals developed eyes with conic surfaces [Cla1975 and Lev1992]. The conic variables developed by Nussbaum [Nus1979 and Nus1998] demonstrate the same principal as the trilobites to add or subtract conic correction to a lens surface to correct for spherical aberration.

In this investigation, the spider eye evolves by random change. The randomization process depends on the computer's random number generator. It is impossible for the computer to generate a true random number [Cha1995]. However, it is possible to generate a random number that is sufficient for a particular the purpose at hand. In order to check whether the computer did generate numbers that were sufficiently random, a test for a billion cycles for the computer used had a random generator accurate to one part in 10,000, which is sufficient with this investigation.

Randomization of the variables in the spider eye proceeded as follows. The first random number chooses the variable to be changed. A second random number picks whether the change is positive or negative. A third random number then chooses the amplitude of the change. Thus, for each cycle, three random numbers are applied.

## 3.0 FURTHER DETAILS OF THE COMPUTER CALCULATION

The Nussbaum optical equations [Nus1979 and Nus1991] are used to calculate the updates for the spider eye at each computer cycle. The method for making the Nussbaum paraxial optical calculations [Nus1991] employs the Gauss matrix , developed in the 1800's. The thick lens nomenclature by Nussbaum [Nus2991] is an excellent representation of the ctenid PM eye lens and is used for the spider eye lens model. Figure 1 (taken from [Nus1991]) shows FFL, R1, R2, T2, N1, N2, N3, M1, M2, BFL. From this list eight variables namely, are R1, R2, M1, M2, N2, N3, T2, BFL are randomized and used in the Nussbaum equations to evolve the spider eye lens. In Figure 1 N1 is a constant because the index of air=1.0000. Items not used for the eye lens calculation are –F, PP1, PP2. T1 also equal to FFL and arbitrarily set to 100 mm for the focal plane from which parallel rays emanate toward the convex lens.

Off-axis rays greater than 5 degrees from the optical axis, is the condition where the Nussbaum optical ray trace equations apply [Nuss1979]. In order to get a representative distribution of rays across the aperture, five rays parallel to the optical axis and spaced at half position across the AP (to obtain radii instead of the diameters), rays at positions 0.16/2AP, 0.35/2AP, 0.50/2AP, 0,65/2AP and 0.85/2AP emanate from the object plane toward the lens . The image after rays are entered into the equations and cycle until SS becomes a maximum of 0.006 mm and f/# becomes between .55 and .90, simultaneously. M1 and M2 are the Nussbaum surface constants that cause the image placement exactly at the BFL and are corrected across the whole aperture. The Nussbaum surface constants, M1 and M2, are determined randomly. Thus, the rays after correction by M1 and M2 image all rays at the BFL. One cycle of calculation consists of these items: 1. Determine the randomized amplitudes for the variables; 2. Run the Nussbaum paraxial equation to find the BFL; 3. Run the Nussbaum ray trace equations with the randomized variables and imaged at the BFL to find value f/# and SS.

A very flexible computer code written in True Basic calculates this mathematical demonstration of evolution. Although the code could probably be more elegant if written by a professional programmer, it is adequate from an optical physicist's point of view.  A difference in the calculation from the model in Figure 1 is that **five** rays emanate from the object plane instead of **three**.  Finally, eight variables are varied randomly during the cycling process. These are: R1, R2, T2, N2, N3, M1, M2, and AP.

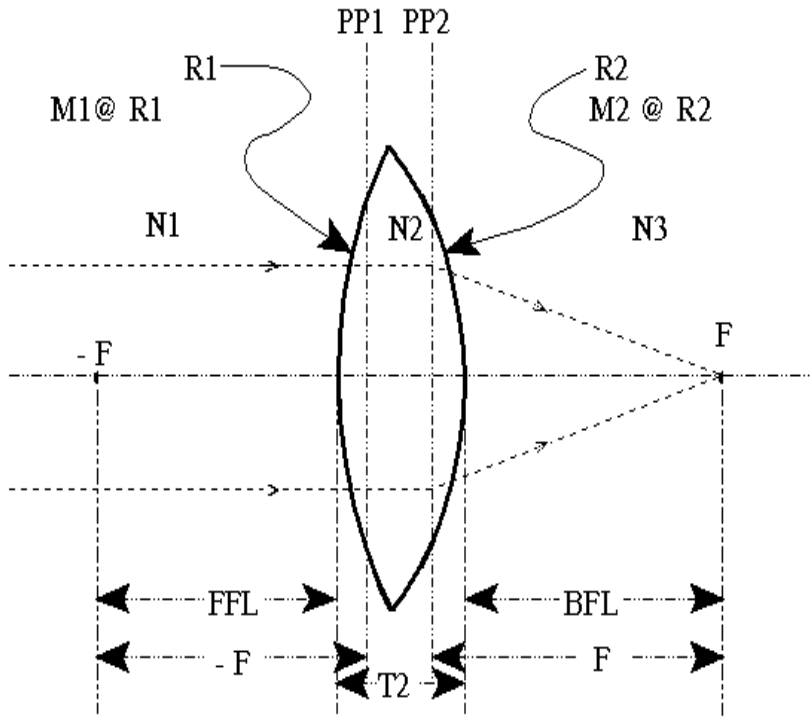

**Figure 1 Thick Lens Model for Ctenid Eye Lens**

|  | Radius [mm] | Diameter [mm] |
|---|---|---|
| First Surface, R1 | .0320 | ------------ |
| Second Surface, R2 | .0236 | ----------- |
| Aperture, AP | ---------- | 0 .625 |

**Table 1 Land Measurements for PM Eye of Ctenid Spider [Lan1992]**

The starting point for the eye lens cycling is what amounts to an eyespot. This eyespot has no surface curvature for front and back surfaces and is flat transparent skin. From this starting point, the computer cycling of the equations begins. During cycling, the lens, the radii, and the lens thickness evolve toward the shape of the ctenid eye lens. Because the computer chooses variables without thought, there must be constraints on the randomly chosen variables. Firstly, the variable must be realizable as no negative lengths. Secondly, indexes of refraction must be naturally realizable for the eye lens material itself as well as the ocular fluid. Thirdly, the space required for the eye lens, that is, from R1 to BFL length must fit within the spider's head. Fourthly, the values for R1, R2, M1, M2 must have a curvature that is not smaller or larger than a limit value. When these four constraints are out of speciation, that cycle is not used.

Two metrics measure the spider's capability to compete in its environment. These are f-number, f/#, and imaged spot size, SS. The specifications for "good eye" are f/# between .55 and .90 and a SS, spot size radius less than 0.006 mm. The f/# is calculated from the value of F/AP. Note that F is the value shown in Figure 1. The SS spot size is calculated from the root mean square of the five rays, parallel to the optical axis, that emanate from the object plane to the image plane.

Ordinarily it takes about 100,000 cycles to meet the appellation of a "good eye." The f/#, as is commonly related to camera operation, is the metric for collecting light; the smaller the f/# the greater ability to collect light. The SS value measures spider eye acuity. Each of the surfaces have Nussbaum conic curvature added, M1 for surface R1 and M2 for surface R2 to correct for spherical aberration. Since the eye lens has air in object space, this index of refraction is fixed at N1=1.0000. The object distance, FFL in Figure 1, is a constant at 100.0000 mm

The values in Table 1 show the radii for the two lens surfaces, R1 and R2, which Land and Barth measured [Lan1991]. These surfaces require correction curvature to correct for spherical aberration. Land and Barth did not include any such corrections in their paper [Lan1991]. The omission could result from the fact that making such measurements is very difficult in excised eyes. A calculation-run made with True Basic program, R1 and R2 values with no M1 or M2 surface correction did not produce a "good eye:"after 30 hours of cycling or about 10,000,000 cycles.

With the same conditions, but using M1 and M2, "good eyes" occurred at a better than normal rate. This result makes sense because R1 and R2 already had the proper values to three decimal places.

## 4. SIMULATED EVOLUTION RESULTS

Hundreds of "good eye" combinations have evolved by the techniques described. A typical set of of "good eye" and variables are depicted in Table 2. The calculations to produce these values are calculated using six decimal places in the True Basic computer program.

| Variable | Amplitude [mm] | Amplitude [numeric] |
|---|---|---|
|  |  |  |
| R1 | .3886 | * |
| R2 | -.3054 | * |
| T1 | 100.0000 | * |
| T2 | .1882 | * |
| BFL | .5120 | * |
| AP | .7601 | * |
| M1 | * | $5.993 \times 10^{-3}$ |
| M2 | * | 1.2453 |
| SS | 1.9422x10-3 | * |
| N1 | * | 1.0000 |
| N2 | * | 1.5166 |
| N3 | * | 1.3541 |
| f/# | * | .8502 |

**Table 2  Typical variables for 300 "Good Eyes"**

Note that the variables in Table 2 show an extremely small value for SS even though the maximum for SS was set to < 0.006 mm. Normally, 100 "good eyes" occur in about 10 million computer cycles. The values for R1 and R2 in Table 2 are similar to the same variables, R1 and R2, in Table 1 but not the same.

| Change to Random Variables | Value for SS [mm] | Value for f/# [numeric] | Effect on SS | Effect to f/# |
|---|---|---|---|---|
|  |  |  |  |  |
| None | 1.9422e-3 | 0.8502 | 0.0 | 0.0 |
| +.001 | 7.7686e-2 | 0.8514 | 40 times larger | 1.12 times larger |
| +.01 | 7.8152e-2 | 0.8617 | 40.2 times larger | 1.01 times larger |
| -.001 | 1.9423e-3 | 0.8490 | 1.00005 times larger. | .001 times smaller |
| -.01 | 1.9850e-3 | 0.8375 | 1.02 times larger | .0145 times smaller |

**Table 3 Effects of Changes to Random Variables Magnitude on SS and f/#**

Table 3 depicts amplitudes for SS and f/# to four decimal places. The "Change to Random Variables" values were made using an auxiliary True Basic program that allows small changes to the variables before calculation. Table 3 also shows two other effects. First, if +.001 and +0.01 is added to all variables SS, and f/#, are no longer within specification for a "good eye". Second, when -.001 is subtracted from the variables, it does not drive the eye out of specification, in fact, the -0.001 produces a better eye. Because the results on line 1 were randomly designed, it shows that is possible that variables be optimized. If the design were made by an optical engineer, the variables would have been optimized more.

## 5. IMPLICATIONS OF THE RESULTS

A number of implications emerge from the ctenid spider eye results presented in section **4**. They are as follows:

- This investigation demonstrates that the ctenid PM eye phenotype does produce a "good eye lens" by random choice of variables, using only those constraints discussed in **3.** This result agrees with Darwin's thinking. The results with the phenotype, although paltry in the light of evolution as a whole, are a first step for mathematically demonstrating evolution. If the phenotype did not evolve, evolution of the genotype would be in doubt.

- There are thousands of combinations for the eight variables used to simulate "good eyes." All these combinations are optically equivalent to the PM eyes measured by Land Barth [Lan1992].

- As previously mentioned, the ideal mathematical proof of evolution is with the genotype. Even if the knowledge did exist for how the genotype codes the animal eye, the knowledge of how to make the calculation is unknown. This type of calculation for the ctenid spider would require parallel computation for the co-evolution of other parts of the spider occurring at the same or nearly the same time as evolution of the eye lens. Thus, to use the genotype, requires super fast computers operating with parallel computation. The vast amount of calculation is similar in magnitude to the work by Castagnoli [Cas 2009] writing about the human brain, in which he shows that the amount of data to be processed is too great for an electronic computer and probably some type of quantum computer would be needed.

- Because the "good eye" develops in spurts when the random variable is in a "good" range, this fact lends credence to Eld [1972] for punctuated equilibrium.

Note: The code written in True Basic is available from the author.